\documentclass[aps,preprint,nofootinbib]{revtex4}
\usepackage{amsfonts}
\usepackage{amsmath}
\usepackage{amssymb}
\usepackage{graphicx}
\usepackage{color}

\begin{document}

\title{Analytic scattering parameters at low energies.}

\author{E.~Z. Liverts}
\affiliation{Racah Institute of Physics, The Hebrew University, Jerusalem 91904,
Israel}


\begin{abstract}
The scattering of two and more particles at low energies is described by the so called effective-range expansion. The leading terms of this expansion are the scattering length and effective range.
The analytic expressions for both of the aforementioned scattering parameters are presented for the inverse-power potential and the Woods-Saxon potential.
A technique for calculating the approximate scattering parameters  is proposed.
Approximate analytic formulas representing the scattering length and effective range are obtained for the Yukawa potential.
The corresponding figures demonstrate a few interesting features of the effective range.
All analytic formulas, both exact and approximate, were verified by comparing with the corresponding results obtained by direct numerical calculations.
Wolfram Mathematica is heavily used.
The presented results can be used with advantage in the fields of nuclear physics,
atomic and molecular physics, quantum chemistry  and many others.
\end{abstract}


\maketitle

\section{Introduction}\label{S0}

The scattering of two particles with short-range interactions at sufficiently low energy is determined by their $S$-wave scattering length $a$.
The energy is sufficiently low if the de Broglie wavelengths of the particles are large compared to the range of the interaction.
The scattering length is important  not only for two-body systems, but also for few-body and many-body systems. If all the constituents of a few-body system have sufficiently low energy, its scattering properties are determined primarily by $a$. A many-body system has properties determined by $a$ if its components have not only sufficiently low energies, but also separations that are large in comparison with the range of interactions.

At sufficiently low energies, the $S$-wave phase shift $\delta_0(k)$ can be expanded in powers of the wave number $k$.
The expansion
\begin{equation}
\label{A0}
k~\cot\delta_0(k)=-\frac{1}{a}+\frac{1}{2}r_ek^2+...
\end{equation}
is called the \textit{effective-range expansion}, and its the first two terms define the $S$-wave scattering length $a$ and the effective range $r_e$ \cite{EBH,FGR}.
These two scattering parameters provide a useful way to parametrize information, e. g., on low-energy nucleon-nucleon scattering. Furthermore, these characteristics may be related to observations other than $nn$-scattering, such as, e. g., deuteron binding energy. Very accurate results can be obtained for the $np$-system by scattering slow neutrons off protons in hydrogen atoms bound in $H_2$ molecules.
Low-energy scattering parameters are widely applicable in the effective field theory (see, e. g., \cite{HAM}) to calculate the properties of nuclear matter and finite nuclei.
For these reasons, a great deal of attention is devoted to the measurements, calculations and understanding of these parameters.

In order to calculate the scattering parameters such as $a$ or $r_e$, one needs to solve the radial Schr$\ddot{\textrm{o}}$dinger equation for $S$-partial waves \cite{LAL,FGR} at zero energy
\begin{equation}
\label{A1} \chi''(r)=\frac{2m}{\hbar^2}V(r)\chi(r),
\end{equation}
under the boundary condition
\begin{equation}
\label{A2}
\chi(0)=0.
\end{equation}
Here $m$ is reduced mass of the considered particles moving in the field $V(r)$ of
a fixed center of force, $\hbar$ is reduced Planck constant.
Using the asymptotic representation
\begin{equation}
\label{A3}
\chi(r)\underset{r\rightarrow \infty}{\simeq} A+B r
\end{equation}
for the solution of Eq.(\ref{A1}) at large enough $r$,
one can find the scattering length $a$ by a simple formula \cite{LAL}:
\begin{equation}
\label{A4}
a=-\frac{A}{B}.
\end{equation}
In turn, the effective range $r_e$ can be calculated by the integral (see, e.g., Ref.\cite{FGR}):
\begin{equation}
\label{B27}
r_e=2\int_0^\infty\left[
\chi_{as}^2(r)-\chi^2(r)
\right] dr,
\end{equation}
where $\chi(r)$ is a physical solution of the Schr$\ddot{\textrm{o}}$dinger equation (\ref{A1}), whereas
\begin{equation}
\label{B27a}
\chi_{as}(r)=\frac{r}{a}-1
\end{equation}
is the specified asymptotic form for $\chi(r)$.

\section{The inverse-power potential}\label{S1}

The inverse-power potential energy is of the form:
\begin{equation}
\label{i1}
V(r)=\frac{\alpha}{r^\beta}.
\end{equation}
In the well-known textbook by Landau and Lifshitz \cite{LAL} one can find exact solutions of this problem for the scattering length $a$ (only!) corresponding to $\alpha>0$ and an integer $\beta>3$.
Here, we derive the exact analytical expressions for both the scattering length $a$ and the effective range $r_e$
representing the more general case of real parameters.

Using the substitutions \cite{LAL}:
\begin{equation}
\label{i2}
r=\left[ \frac{2\sqrt{\lambda}}{(\beta-2)x}\right]^{\frac{2}{\beta-2}} ,~~
\chi(r)=\varphi(x)\sqrt{r},
\end{equation}
\begin{equation}
\label{i4}
\lambda=\frac{2m\alpha}{\hbar^2},~~~\nu=\frac{1}{\beta-2}.
\end{equation}
we can cast the radial Schr$\ddot{\textrm{o}}$dinger equation (\ref{A1}) with potential (\ref{i1})
to a differential equation having an analytic solution $\varphi(x)$.
However, the form of this equation, and hence its solution, depends on the properties of the parameters included in this equation.
From the choice of parameters satisfying the conditions $\alpha>0$ and $\beta>3$, it follows that
the new parameters satisfy the conditions $\lambda>0$ and $0\leq\nu<1$.
For such parameters Eq.(\ref{A1}) is reduced to the differential equation
\begin{equation}
\label{i3}
x^2\varphi''(x)+x\varphi'(x)-(x^2+\nu^2)\varphi(x)=0
\end{equation}
for the function $\varphi(x)$.
The solutions of Eq.(\ref{i3}) are the modified Bessel function $I_{\nu}(x)$ and $K_{\nu}(x)$ of the first and the second kind, respectively \cite{ABS}.
It follows from the definitions (\ref{i2}) that under the conditions mentioned above, the new variable $x$ tends to infinity as $r\rightarrow 0$ .
The asymptotic representations for the functions $I_{\nu}(x)$ show the exponential divergence \cite{ABS} for $x\rightarrow\infty$.
Thus, only the function $K_\nu(x)$ can provide the boundary condition (\ref{A2}).
This yields
\begin{equation}
\label{i5}
\chi(r)= C\sqrt{r}K_\nu(x),
\end{equation}
where $C$ is arbitrary constant and
\begin{equation}
\label{i6}
x=2\nu\sqrt{\lambda}r^{-\frac{1}{2\nu}}.
\end{equation}
Using the well-known formula \cite{ABS} for the series expansion of $K_\nu(x)$, one obtains for the asymptotic representation of the function $\chi(r)$:
\begin{equation}
\label{i7}
\chi(r)\underset{r\rightarrow \infty}{\sim} \frac{C}{2} \left[
\left(\nu \sqrt{\lambda} \right)^{\nu} \Gamma(-\nu)+
\left(\nu \sqrt{\lambda} \right)^{-\nu} \Gamma(\nu) r \right] .
\end{equation}
It should be emphasized that the asymptotic representation (\ref{i7}) is valid only for $0\leq\nu<1$ (and hence for $\beta>3$).
According to definition (\ref{A4}), the ratio of the coefficients in the RHS of Eq.(\ref{i7}) yields the scattering length in the form
\begin{equation}
\label{i8}
a=-\left( \nu \sqrt{\lambda}\right)^{2\nu}\frac{\Gamma(-\nu)}{\Gamma(\nu)}.
\end{equation}
This result coincides completely, of course, with expression for integer $\beta > 3$,
presented in Ref. \cite{LAL}.

To calculate the effective range $r_e$ using Eq.(\ref{B27}), we need a reduced radial function $\chi(r)$ with asymptotic behavior (\ref{B27a}).
This can be achieved putting
\begin{equation}
\label{i9}
C=-\frac{2}{\left(\nu\sqrt{\lambda} \right)^\nu \Gamma(-\nu) }
\end{equation}
in the asymptotic representation (\ref{i7}).

Using definitions (\ref{i6}) and (\ref{i8}),
the first indefinite integral in the representation (\ref{B27}) can be written in the form:
\begin{eqnarray}
\label{i10}
\omega_{as}(x)\equiv \int \chi_{as}^2(r)dr=r-\frac{r^2}{a}+\frac{r^3}{3a^2}=
~~~~~~~~~~~~~~~~~~~~~~~~~~~~~~~~~~~~~~~~~~~~~~~~~~~~~~~\nonumber~\\
\left( \frac{2\nu\sqrt{\lambda}}{x}\right)^{2\nu}+
\left( \frac{4\nu\sqrt{\lambda}}{x^2}\right)^{2\nu}\frac{\Gamma(\nu)}{\Gamma(-\nu)}+
\frac{1}{3}\left( \frac{8\nu\sqrt{\lambda}}{x^3}\right)^{2\nu}\frac{\Gamma^2(\nu)}{\Gamma^2(-\nu)}.
~~~~~~~~~~~~~~~~~~~
\end{eqnarray}
It is seen that $\omega_{as}(\infty)=0$  (for $r\rightarrow 0$), whereas  $\omega_{as}(x\rightarrow0)$ behaves as the divergent expression in the second line of Eq.(\ref{i10}), as $r\rightarrow \infty$.

Using the solution (\ref{i5}) with factor $C$ defined by Eq.(\ref{i9}), the second integral in the RHS of Eq.(\ref{B27}) can be presented in the form:
\begin{equation}
\label{i11}
\int_0^\infty \chi^2(r)dr=\xi
\int_0^\infty x^{-4\nu-1}K^2_\nu(x)dx,
\end{equation}
with
\begin{equation}
\label{i12}
\xi=\frac{8\nu(4\nu\sqrt{\lambda})^{2\nu}}{\Gamma^2(-\nu)}.
\end{equation}
The corresponding indefinite integral can be expressed in terms of the generalized hypergeometric functions $_2F_3(...)$ as follows:
\begin{eqnarray}
\label{i13}
\Omega(x)\equiv\int x^{-4\nu-1}K_\nu^2(x)dx=
~~~~~~~~~~~~~~~~~~~~~~~~~~~~~~~~~~~~~~~~~~~~~~~~~~~~~~~~~~~~~~~~~~\nonumber~\\
=-\frac{\Gamma^2(-\nu)}{8\nu}\left( \frac{1}{2x}\right)^{2\nu}
~_2F_3\left(-\nu,\frac{1}{2}+\nu;1-\nu,1+\nu,1+2\nu;x^2 \right)+
~\nonumber~\\
+\frac{\pi \csc(\pi\nu)}{8\nu^2}\left( \frac{1}{x^2}\right)^{2\nu}~_2F_3\left( \frac{1}{2},-2\nu;1-2\nu,1-\nu,1+\nu;x^2\right) -
~~~~~~~~~~~\nonumber~\\
-\frac{\Gamma^2(\nu)}{24\nu}\left(\frac{2}{x^3} \right)^{2\nu}
~_2F_3\left(\frac{1}{2}-\nu,-3\nu;1-3\nu,1-2\nu,1-\nu;x^2 \right).
~~~~~~~~~~~~~~~~~
\end{eqnarray}
First of all, one should analyse the behavior of the function $\Omega(x)$  as $x\rightarrow 0$.
It is easy to verify that the leading terms of this divergent (as $x\rightarrow 0 $) function are obtained if we restrict ourselves to the first terms of the hypergeometric series equal to one.
In other words, for small enough $x$ we have:
\begin{equation}
\label{i14}
\Omega(x)\underset{x\rightarrow 0}{\simeq}
-\frac{\Gamma^2(-\nu)}{8\nu}\left( \frac{1}{2x}\right)^{2\nu}
+\frac{\pi \csc(\pi\nu)}{8\nu^2}\left( \frac{1}{x^2}\right)^{2\nu}
-\frac{\Gamma^2(\nu)}{24\nu}\left(\frac{2}{x^3} \right)^{2\nu}.
\end{equation}
Multiplying the RHS of Eq.(\ref{i14}) by the factor $\xi$, one obtains $-\omega_{as}(x)$.
This means, that the divergent parts of the integrals in (\ref{B27}) as $r\rightarrow\infty$  (hence, as $x\rightarrow0$) cancel each other out.
Thus, one obtains:
\begin{equation}
\label{i16}
r_e=-2\xi \lim_{x\rightarrow\infty}\Omega(x).
\end{equation}
To calculate $\Omega(\infty)$, let us write down the asymptotic ( $x\rightarrow \infty$)
expressions for the generalized hypergeometric functions presenting in Eq.(\ref{i13}). In particular, for $x>0$ and $\nu>0$ we have:
\begin{eqnarray}
\label{i18}
~_2F_3\left(-\nu,\frac{1}{2}+\nu;1-\nu,1+\nu,1+2\nu;x^2 \right)\underset{x\rightarrow \infty}{\simeq}
\frac{\pi\nu\Gamma\left( \frac{1}{2}+2\nu\right)(-x^2)^\nu }
{\sin(\pi\nu)\Gamma\left(\frac{1}{2}+\nu \right)\Gamma(1+3\nu) }-
\frac{\nu\Gamma^2(1+\nu)e^{2x}}{\pi 2^{1-2\nu}x^{2+2\nu}},
~~~~~~~~~~\nonumber~\\
~_2F_3\left(\frac{1}{2},-2\nu;1-2\nu,1-\nu,1+\nu;x^2 \right)\underset{x\rightarrow \infty} \simeq
\frac{\Gamma(1-2\nu)\Gamma(1-\nu)\Gamma\left(\frac{1}{2}+2\nu \right)
](-x^2)^{2\nu}}{\sqrt{\pi}\Gamma(1+3\nu)}-
\frac{\nu^2e^{2x}}{\sin(\pi\nu)x^2},
~~~~~~\nonumber~\\
~_2F_3\left(\frac{1}2{-\nu,-3\nu;1-3\nu,1-2\nu,1-\nu;x^2} \right) \simeq
~~~~~~~~~~~~~~~~~~~~~~~~~~~~~~~~~~~~~~~~~~~~~~~~~~~~~~~~~~~~~~~~~~~~~~\nonumber~\\
\underset{x\rightarrow \infty}\simeq\frac{\Gamma^2(1-\nu)}{4^\nu}\left[
\frac{\Gamma(1-3\nu)\Gamma\left(\frac{1}{2}+2\nu \right)(-x^2)^{3\nu} }
{\sqrt{\pi}\Gamma(1+\nu)\Gamma(1+2\nu)}-\frac{3\nu e^{2x}}{2\pi x^{2-2\nu}}
\right].~~~~~~~~~
\end{eqnarray}
Substitution of the asymptotic expressions (\ref{i18}) into the RHS of Eq.(\ref{i13})
yields finally:
\begin{equation}
\label{i19}
\Omega(\infty)=\frac{\pi^{3/2}\cos(2\pi\nu)\Gamma(-\nu)\Gamma\left(\frac{1}{2}+2\nu \right) }{2\left[\cos(\pi\nu)-\cos(5\pi\nu) \right]\Gamma(1+2\nu)\Gamma(1+3\nu) }.
\end{equation}
Note, that the terms with $\exp(2x)$ are cancelled from the final expression.

\begin{figure}
\caption{Scattering length $a$ for the inverse-power potential of the form $V(r)=\alpha/r^\beta$. Parameter $\alpha$ is expressed in the units of $\hbar^2/2m$ ($\beta>3$).}
\includegraphics[width=6.0in]{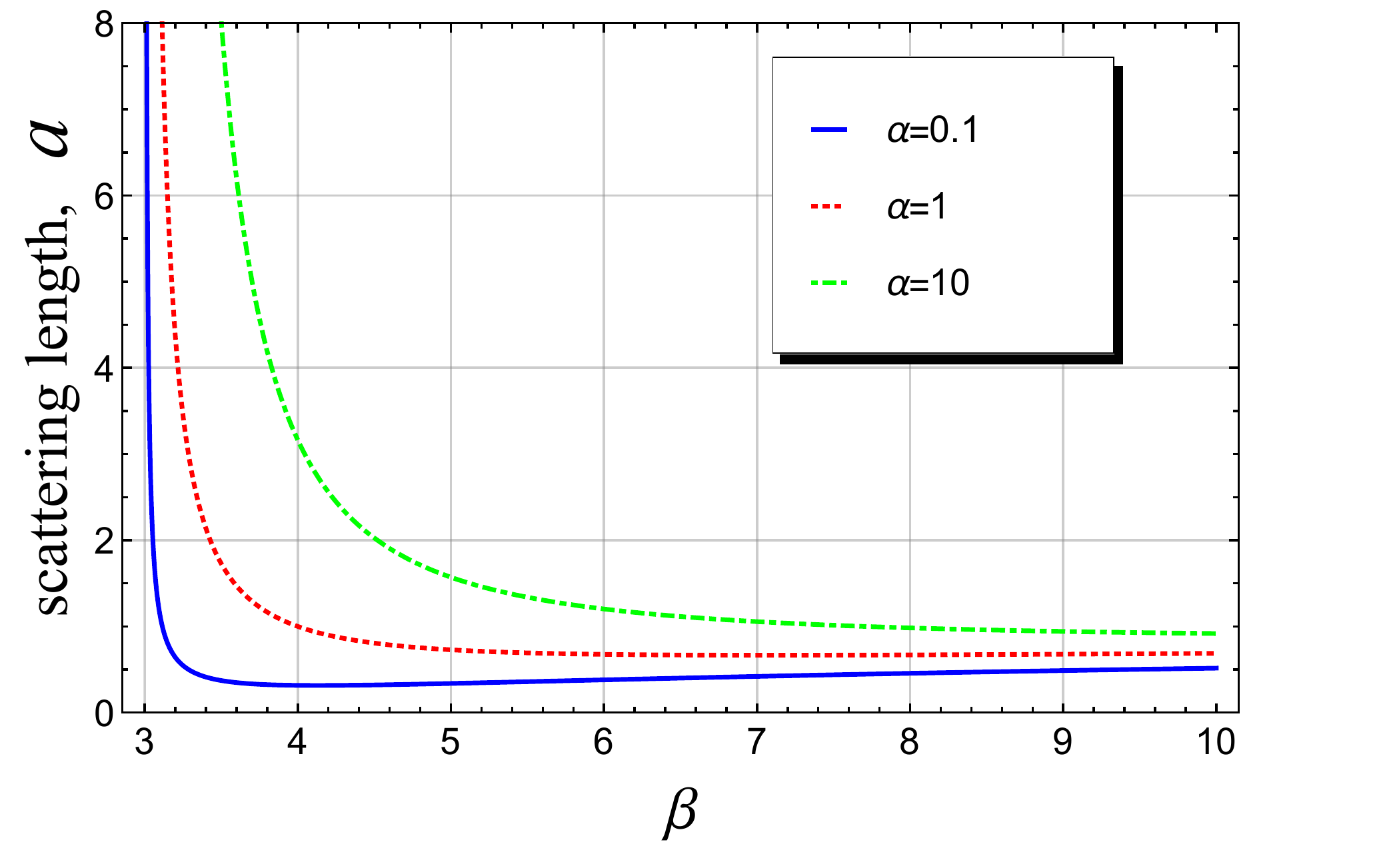}
\label{F1}
\end{figure}

\begin{figure}
\caption{Effective range $r_e$ for the inverse-power potential of the form $V(r)=\alpha/r^\beta$ with $3<\beta<6$. Parameter $\alpha$ is expressed in the units of $\hbar^2/2m$.}
\includegraphics[width=6.0in]{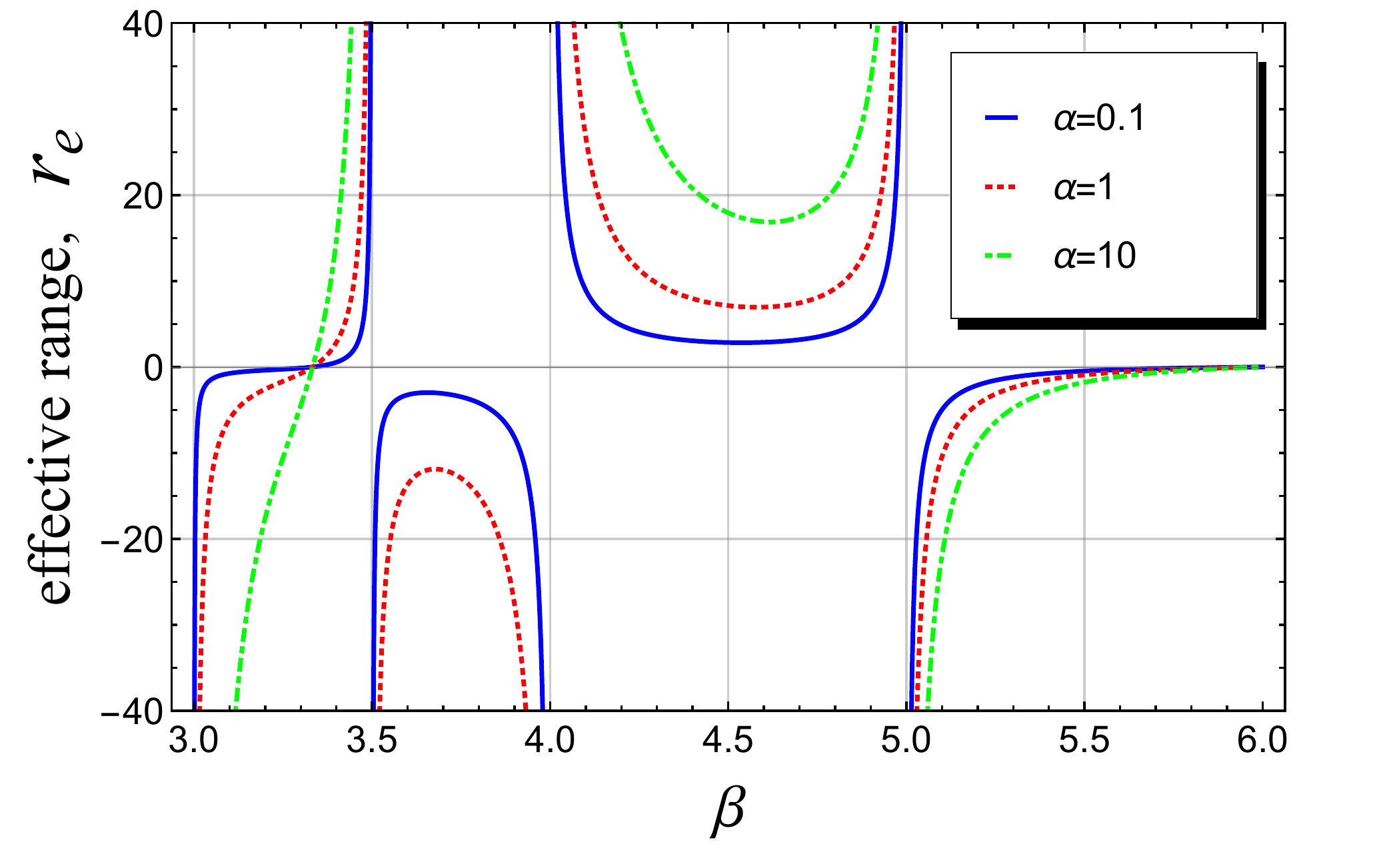}
\label{F2}
\end{figure}

\begin{figure}
\caption{Effective range $r_e$ for the inverse-power potential of the form $V(r)=\alpha/r^\beta$ with $\beta>6$. Parameter $\alpha$ is expressed in the units of $\hbar^2/2m$.}
\includegraphics[width=6.0in]{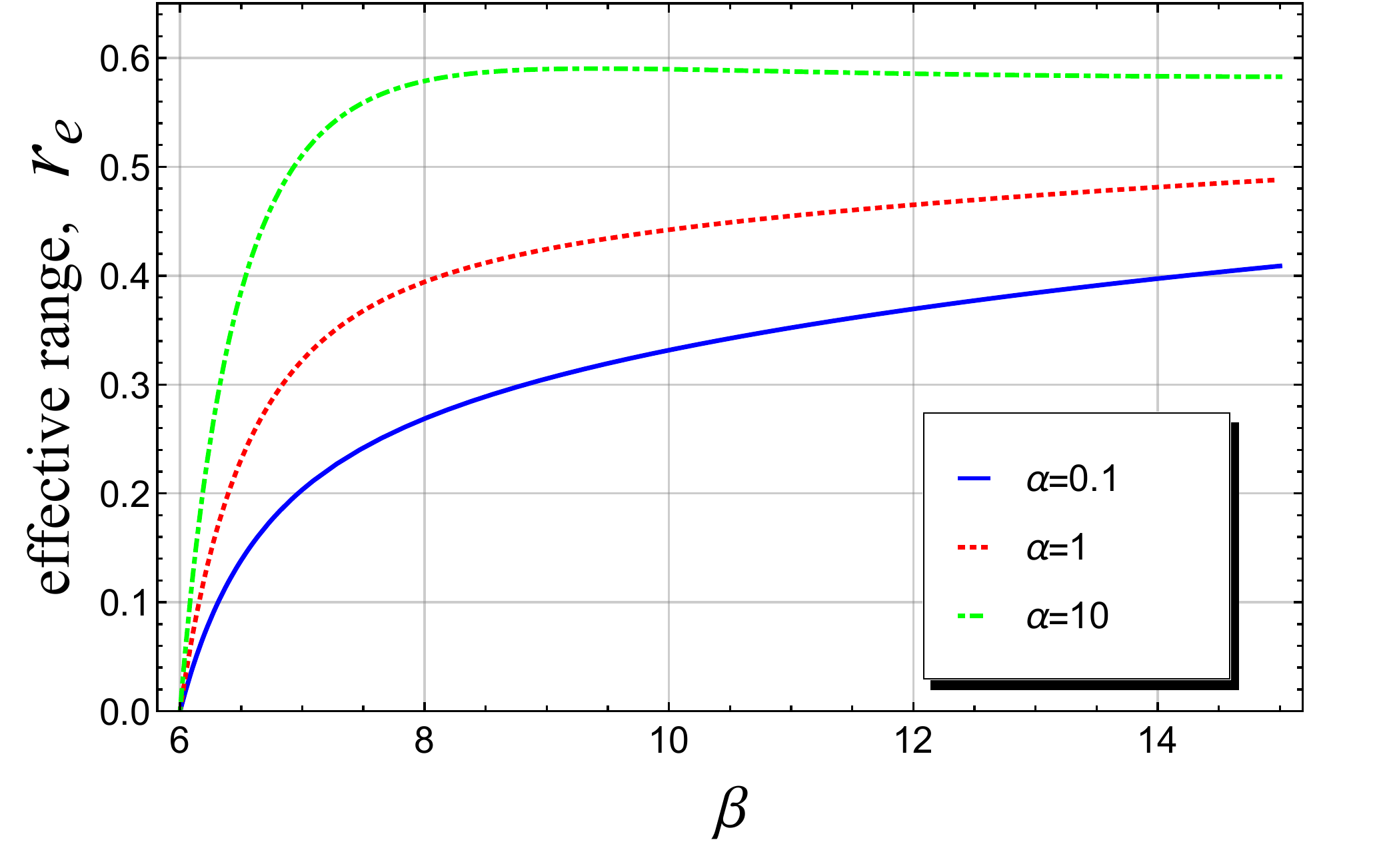}
\label{F3}
\end{figure}

Performing some manipulations, one obtains the following simple expression for the effective range according to Eqs.(\ref{i12}), (\ref{i16}) and (\ref{i19}):
\begin{equation}
\label{i20}
r_e=\frac{4\pi^{3/2}\left(4\nu \right)^{2\nu}\cos(2\pi\nu)
\Gamma\left(\frac{1}{2}+2\nu \right) \lambda^\nu }
{\left[ \cos(5\pi\nu)-\cos(\pi\nu)\right]\Gamma(-\nu)\Gamma(2\nu)\Gamma(1+3\nu) }.
\end{equation}
It is useful to present the two limit values for the scattering length and effective range represented by Eqs.(\ref{i8}) and (\ref{i20}), respectively:
\begin{equation}
\label{i21}
\lim_{\beta\rightarrow\infty}a=1,~~~~~~\lim_{\beta\rightarrow\infty}r_e=\frac{2}{3}.
\end{equation}
The plots of the scattering length $a$ and the effective range $r_e$ for the inverse-power potential (\ref{i1}) with parameters $\beta>3$ and $\alpha>0$ are presented in Fig.\ref{F1} and Figs.\ref{F2}-\ref{F3}. The plot of $a$ as a function of $\beta>3$ is drawn using the expressions (\ref{i8}) and (\ref{i4}) for $\alpha=0.1, 1, 10$ in the units of $\hbar^2/2m$. The plots of $r_e$ as a function of $\beta$ are drawn using the expressions (\ref{i20}) and (\ref{i4}) for the same values of
$\alpha$ as in Fig.\ref{F1}.
Because of the different scales, the graphs for $3<\beta<6$ and $\beta>6$ are presented on individual figures, which demonstrate a few interesting features of the effective range.
First, there are two points of zero effective range: $\beta=10/3$ and $\beta=6$. This follows directly from definition (\ref{i20}) for $\nu=3/4$ and $\nu=1/4$, respectively. Second, there ara three $\beta$-intervals for negative $r_e$:[0,10/3],[3.5,4],[5,6], and three $\beta$-intervals for positive $r_e$:[10/3,3.5],[4,5],[6,$\infty$].
Third, there are four $\beta$-points of (infinite) discontinuity of the second kind : $\beta=3,3.5,4,5$.

\section{The Woods-Saxon potential}\label{S2}

The standard representation for the Woods-Saxon potential is of the form:
\begin{equation}
\label{w1}
V(r)=-\frac{U_0}{1+\exp(\frac{r-R}{\eta})}.~~~~~~~~~~~~~~[U_0>0,~R>0,~\eta>0]
\end{equation}
The analytical solutions of the Schr$\ddot{\textrm{o}}$dinger equation with generalized Woods-Saxon potential were presented in Ref. \cite{CBB}.
An assumption about using these results for zero energy can appear in mind.
However, this idea raises reasonable doubts, if only because the authors used the so-called Nikiforov-Uvarov method which basically is an approximate one.
Instead, we obtain the exact solution of the Schr$\ddot{\textrm{o}}$dinger equation in question.

To find a general solution of the Schr$\ddot{\textrm{o}}$dinger equation (\ref{A1}) with the
Woods-Saxon potential (\ref{w1}) let us introduce a new variable
\begin{equation}
\label{w2}
x=\exp\left( \frac{r-R}{\eta}\right),
\end{equation}
and a new parameter
\begin{equation}
\label{w2a}
\alpha=\frac{\eta}{\hbar}\sqrt{2 m U_0}.
\end{equation}
Instead of Eq.(\ref{A1}), we then obtain the following differential equation for the function $\varphi(x)\equiv \chi(r)$:
\begin{equation}
\label{w2b}
x^2\varphi''(x)+x\varphi'(x)+\frac{\alpha^2}{1+x}\varphi(x)=0.
\end{equation}
Wolfram Mathematica yields the general solution of Eq.(\ref{w2b}) in the form
\begin{equation}
\label{w3}
\varphi(x)=N_1F(\alpha,x)+N_2F(-\alpha,x).
\end{equation}
We introduced notation
\begin{equation}
\label{w5} F(\alpha,x)=x^{i\alpha}~_2F_1(i\alpha,i\alpha;1+2i\alpha;-x),
\end{equation}
where $_2F_1(a,b;c;z)$ is the Gauss hypergeometric function, and $i=\sqrt{-1}$ is the imaginary unit.

Employing the boundary condition (\ref{A2}), one can present the solution of
Eq.(\ref{A1}) for the Woods-Saxon potential in the form:
\begin{equation}
\label{w4} \varphi(x)=N\left[ F(-\alpha,x_0)F(\alpha,x)-F(\alpha,x_0)F(-\alpha,x) \right],
\end{equation}
where
\begin{equation}
\label{w6} x_0\equiv \lim_{r\rightarrow0}x=\exp(-R/\eta),
\end{equation}
and $N$ is arbitrary constant.

In order to study the asymptotic behavior of the function (\ref{w4}), one can
use the representation (15.3.13) from Handbook \cite{ABS}. For the considered case, this
yields
\begin{eqnarray}
\label{w7} x^b~_2F_1(b,b;1+2b;-x)=\frac{2\Gamma(2b)}{\Gamma^2(b)}\times
~~~~~~~~~~~~~~~~~~~~~~~~~~~~~~~~~~~\nonumber~\\
\times\sum_{n=0}^\infty\frac{\Gamma(n+b)\Gamma(n-b)}{\Gamma(b)\Gamma(-b)(n!)^2}
(-x)^{-n}\left[ \ln(x)+2\psi(n+1)-\psi(b+n)-\psi(b+1-n) \right] ,~~~~~|x|\geq1.~~~~
\end{eqnarray}
where, $\psi(z)=\Gamma'(z)/\Gamma(z)$ is the logarithmic derivative of the Euler gamma function (digamma function). Eq.(\ref{w7}) shows, that the asymptotic ($x\rightarrow \infty$) behavior of its RHS is determined by the term with $n=0$. Hence, for large enough $x$ one has:
\begin{equation}
\label{w8}
 x^b~_2F_1(b,b;1+2b;-x)\underset{x\rightarrow \infty}{\simeq} \frac{2\Gamma(2b)}{\Gamma^2(b)}
\left[ \ln(x)-2\gamma-2\psi(b)-\frac{1}{b} \right],
\end{equation}
where $\gamma$ is the Euler`s constant. Using expression (\ref{w8})
with $x$ defined by Eq.(\ref{w2}), we obtain for the asymptotic representation (\ref{A3}):
\begin{eqnarray}
\label{w9}
\chi(r)\underset{r\rightarrow \infty}{\simeq}\frac{2N}{\alpha}\left\{
\frac{F(\alpha,x_0)\Gamma(-2i\alpha)\left[ i+2\alpha
\left(\gamma+\psi(-i\alpha)\right)-\alpha(r-R)/\eta\right]}{\Gamma^2(-i\alpha)}~+
\right.
~~~~~~~~~~~~~~~~~~~~~~~~~~\nonumber~\\
\left. \frac{F(-\alpha,x_0)\Gamma(2i\alpha)\left[ i-2\alpha
\left(\gamma+\psi(i\alpha)\right)+\alpha(r-R)/\eta\right]}{\Gamma^2(i\alpha)}
\right\}.~~~~~~~~~~~~~~~~~~~~~~
\end{eqnarray}
The latter equation enables one to derive the analytic expression for the scattering
length according to Eqs.(\ref{A3})-(\ref{A4}):
\begin{equation}
\label{w10}
a=R+2\eta\left[\gamma+Re~\psi(i\alpha)-\frac{i \pi}{2}\coth(\pi \alpha)\
\frac{F(\alpha,x_0)\Gamma^2(i\alpha)\Gamma(-2i\alpha)+
F(-\alpha,x_0)\Gamma^2(-i\alpha)\Gamma(2i\alpha)}
{F(\alpha,x_0)\Gamma^2(i\alpha)\Gamma(-2i\alpha)-
F(-\alpha,x_0)\Gamma^2(-i\alpha)\Gamma(2i\alpha)}\right].~~~
\end{equation}
Deducing this formula, we used the following properties of the digamma function
\cite{ABS}:

$~~~~~~~\textrm{Re}~\psi(i \alpha)=\textrm{Re}~\psi(-i \alpha)$,

$~~~~~~~\textrm{Im}~\psi(i \alpha)=-\textrm{Im}~\psi(-i \alpha)=\frac{1}{2\alpha}+\frac{\pi}{2}\coth(\pi
\alpha)$.

Note, that the denominator in the  expression (\ref{w10}) can be reduced to the more
good-looking form:
\begin{equation}
\label{w11} F(\alpha,x_0)\Gamma^2(i\alpha)\Gamma(-2i\alpha)-
F(-\alpha,x_0)\Gamma^2(-i\alpha)\Gamma(2i\alpha)=-\frac{i \pi^2}{\alpha \sinh^2(\pi
\alpha)}~_2F_1\left(-i \alpha,i \alpha;1;-\frac{1}{x_0}\right).
\end{equation}
Thus, one obtains the alternate representation for the scattering length:
\begin{equation}
\label{w12} a=R+2\eta\left[\gamma+\textrm{Re}~\psi(i\alpha)+\frac{\alpha}{4\pi}\sinh(2\pi
\alpha)\ \frac{F(\alpha,x_0)\Gamma^2(i\alpha)\Gamma(-2i\alpha)+
F(-\alpha,x_0)\Gamma^2(-i\alpha)\Gamma(2i\alpha)} {~_2F_1\left(
-i\alpha,i\alpha;1;-\frac{1}{x_0} \right)}\right].~~~
\end{equation}
In order to calculate the effective range $r_e$ corresponding to the Woods-Saxon potential (\ref{w1}), it is necessary to obtain the wave function (\ref{w4}) with asymptotic behavior (\ref{B27a}) (see, Eq.(\ref{B27})).
This can be realized by means of special choice of the constant $N$ in Eq.(\ref{w4}).
The required expression for $N$ can be determined from the condition that the coefficient for $r^0$ in the RHS of Eq.(\ref{w9}) must be equal to $(-1)$. This yields:
\begin{eqnarray}
\label{w13} N=\frac{\pi^2}{2\alpha \sinh^2(\pi \alpha)}\left\{
F(-\alpha,x_0)\Gamma^2(-i\alpha)\Gamma(2i\alpha) \left[
R\alpha/\eta-i+2\alpha\left(\gamma+\psi(i\alpha)\right) \right]- \right.
~~~~~~~~~~~~~\nonumber~\\
\left.- F(\alpha,x_0)\Gamma^2(i\alpha)\Gamma(-2i\alpha) \left[
R\alpha/\eta+i+2\alpha\left(\gamma+\psi(-i\alpha)\right) \right]
\right\}^{-1}.~~~~~~~~~~~~~~
\end{eqnarray}

Thus, according to definition (\ref{B27}) for the effective range, one can write down:
\begin{eqnarray}\label{w14}
r_e\equiv2\int_0^\infty \left[\left( \frac{r}{a}-1 \right)^2-\chi^2(r)  \right]dr=
~~~~~~~~~~~~~~~~~~~~~~~~~~~~~~~~~~~~~~~~~~~~~~~~~~~~~\nonumber~\\
=2\eta\left\lbrace
I_1-N^2\left[
F^2(-\alpha,x_0)J_1+F^2(\alpha,x_0)J_2-2F(\alpha,x_0)F(-\alpha,x_0)J_3
\right]
\right\rbrace,
\end{eqnarray}
where
\begin{equation}
\label{w15}
I_1=\int_{x_0}^\infty\left(\frac{R+\eta \ln x}{a}-1 \right) ^2\frac{dx}{x},
\end{equation}
\begin{equation}
\label{w16}
J_1=\int_{x_0}^\infty F^2(\alpha,x)
\frac{dx}{x},
\end{equation}
\begin{equation}
\label{w17}
J_2=\int_{x_0}^\infty F^2(-\alpha,x)
\frac{dx}{x},
\end{equation}
\begin{equation}
\label{w18}
J_3=\int_{x_0}^\infty F(\alpha,x)F(-\alpha,x)
\frac{dx}{x}.
\end{equation}
Parameters $a,~N$ and $x_0$ are defined by Eqs.(\ref{w12}), (\ref{w13}) and (\ref{w6}), respectively.

The integral (\ref{w15}) corresponding to the asymptotic function (\ref{B27a}) can be presented in the form:
\begin{equation}
\label{w19}
2\eta I_1=\zeta+\frac{2\eta}{a^2}\lim_{x\rightarrow\infty}
\left[\frac{\eta^2}{3}\ln^3 x+\eta(R-a)\ln^2 x +(R-a)^2 \ln x\right] ,
\end{equation}
with
\begin{equation}
\label{w20}
\zeta=2R\left(1-\frac{R}{a}+\frac{R^2}{3a^2} \right).
\end{equation}
Unfortunately, no one of integrals (\ref{w16})-(\ref{w18}) can be taken in the explicit form.
However, it is still possible to derive an analytic expression for the effective range (\ref{w14}) by the use of the following method.

First of all, it is important to note that the parameter $x_0<1$ by definition (\ref{w6}).
Furthermore, for the considered in what follows example of the optical-model calculations, this parameter is very small (for this case $x_0<0.0012$).
This enables us to split the range of integration $[x_0,\infty)$ in Eqs.(\ref{w16}-\ref{w18}) into two parts:
$[x_0,1]$ and $[1,\infty)$.
Accordingly, it will be wise to use the asymptotic representation (\ref{w7}) for integration over the second range, while for the first range it is natural to use a power series representing the  Gauss hypergeometric function included in Eq.(\ref{w5}).

It can be shown that the divergent logarithmic part represented in Eq.(\ref{w19}) is exactly eliminated by the proper logarithmic terms that arise when taking the upper limit ($\infty$) of integration  in Eqs.(\ref{w16})-(\ref{w18}).
Note that the latter limit of integration represents simultaneously the upper limit for the second range of integration. For the lower limit of integration (equals $1$) in the second range, one obtains:
\begin{eqnarray}
\label{e1}
J_1^{(L)}\equiv\left.\int F_{large}^2(\alpha,x)\frac{dx}{x}\right|_{x=1}=
~~~~~~~~~~~~~~~~~~~~~~~~~~~~~~~~~~~~~~~~~~~~~~~~~~~~~~~~~~~~~~~~~~~~~~\nonumber~\\
=-\frac{4\Gamma^2(2i\alpha)}{\Gamma^6(i\alpha)\Gamma^2(-i\alpha)}
\underset{(n+m>0)}{\sum_{n=0}^\infty\sum_{m=0}^\infty}\frac{\Gamma(n+i\alpha)
\Gamma(n-i\alpha)\Gamma(m+i\alpha)\Gamma(m-i\alpha)}{(n!)^2(m!)^2(m+n)^3(-1)^{n+m}}
\left\lbrace
2+(m+n)^2
\left[ 2\psi(n+1)-
\right.
\right.
~\nonumber~\\
\left.
\left.
-\psi(1+i\alpha-n)-\psi(n+i\alpha)\right]
\left[ 2\psi(m+1)-\psi(1+i\alpha-m)-\psi(m+i\alpha)\right]+(m+n)\times
\right.
~~~~~~~~~~~~~\nonumber~\\
\left.
\times\left[
2\psi(n+1)+2\psi(m+1)-\psi(1+i\alpha-n)-\psi(1+i\alpha-m)-\psi(n+i\alpha)-\psi(m+i\alpha)\right]
\right\rbrace.~~~~~~~~
\end{eqnarray}
This result was obtained by taking the indefinite integral corresponding to the RHS of Eq.(\ref{w16}) and then replacing $x$ by $1$ in the resulting expression.
It is important that just the Eq.(\ref{w7}) was used to represent the Gaussian hypergeometric functions
included in the corresponding integrands through the representation (\ref{w5}).

It is easy to verify that $J_2^{(L)}=\left(J_1^{(L)}\right)^*$ where the asterisk denotes complex conjugation.
Using again the asymptotic representation (\ref{w7}) one obtains for the indefinite integral in the RHS of Eq.(\ref{w18}) taken at $x=1$:
\begin{eqnarray}
\label{e2}
J_3^{(L)}\equiv\left.\int F_{large}(\alpha,x)F_{large}(-\alpha,x)\frac{dx}{x}\right|_{x=1}=
~~~~~~~~~~~~~~~~~~~~~~~~~~~~~~~~~~~~~~~~~~~~~~~~~~~~~~~~\nonumber~\\
=-\frac{\alpha^3\sinh^3(\pi\alpha)}{\pi^3\cosh(\pi\alpha)}
\underset{(n+m>0)}{\sum_{n=0}^\infty\sum_{m=0}^\infty}
\frac{\Gamma(n+i\alpha)\Gamma(n-i\alpha)\Gamma(m+i\alpha)\Gamma(m-i\alpha)}{(n!)^2(m!)^2(m+n)^3(-1)^{n+m}}
\left\lbrace 2+(m+n)^2
\left[ 2\psi(n+1)-
\right.\right.
~~~~~\nonumber~\\
\left.\left.
-\psi(1+i\alpha-n)-\psi(n+i\alpha)\right]
\left[ 2\psi(m+1)-\psi(1-i\alpha-m)-\psi(m-i\alpha)\right]+(m+n)\times
\right.
~~~~~~~~~~~~~~\nonumber~\\
\left.
\times\left[
2\psi(n+1)+2\psi(m+1)-\psi(1+i\alpha-n)-\psi(1-i\alpha-m)-\psi(n+i\alpha)-\psi(m-i\alpha)\right]
\right\rbrace.~~~~~~~~~~
\end{eqnarray}
On the other hand, no problem to take the indefinite integrals representing the RHS of Eqs.(\ref{w16})-(\ref{w18}) making use a power series representation for the Gauss hypergeometric functions. This yields:
\begin{eqnarray}
\label{e3}
J_1^{(S)}(x)\equiv\left( J_2^{(S)}\right)^*(x)=
~~~~~~~~~~~~~~~~~~~~~~~~~~~~~~~~~~~~~~~~~~~~~~~~~~~~~~~~~~~~~~~~~~~~~~~~~~~~~~~~~\nonumber~\\
=\frac{x^{2i\alpha}\Gamma^2(1+2i\alpha)}{\Gamma^4(i\alpha)}
\sum_{n=0}^\infty\sum_{m=0}^\infty\frac{\Gamma^2(n+i\alpha)\Gamma^2(m+i\alpha)(-x)^{n+m}}
{\Gamma(1+n+2i\alpha)\Gamma(1+m+2i\alpha)n!m!(n+m+2i\alpha)},~~~~~~~~~~~~
\end{eqnarray}
\begin{equation}
\label{e4}
J_3^{(S)}(x)=\ln x+\frac{\alpha^3\tanh(\pi\alpha)}{\pi}
\underset{(n+m>0)}{\sum_{n=0}^\infty\sum_{m=0}^\infty}\frac{\Gamma^2(n+i\alpha)
\Gamma^2(m-i\alpha)(-x)^{n+m}}{\Gamma(1+n+2i\alpha)\Gamma(1+m-2i\alpha)n!m!(n+m)}.~~~~~
\end{equation}
Using representations (\ref{e1})-(\ref{e4}), one can rewrite Eq.(\ref{w14}) in the final analytic form:
\begin{eqnarray}
\label{e5}
r_e=\zeta+2\eta N^2\left[ F^2(-\alpha,x_0)
\left(J_1^{(S)}(x_0)+J_1^{(L)}-J_1^{(S)}(1)\right)+
\right.
~~~~~~~~~~~~~~~~~~~~~~~~~~~~~~~~~~~~~~~~~~~~~~~~~~~~\nonumber~\\
\left.
+F^2(\alpha,x_0)\left(J_2^{(S)}(x_0)+J_2^{(L)}-J_2^{(S)}(1)\right)
-2F(\alpha,x_0)F(-\alpha,x_0)
\left(J_3^{(S)}(x_0)+J_3^{(L)}-J_3^{(S)}(1)\right)\right], ~~~~~~
\end{eqnarray}
where $N$ and $\zeta$ are defined by Eqs.(\ref{w13}) and (\ref{w20}), respectively.

Parameters of the Woods-Saxon potential, presented in Ref. \cite{PER,MJ1} for the optical-model calculations (see, also Ref. \cite{MJ2}), have been adopted, as an example.
These are: $R=1.285 A^{1/3}~\textrm{fm};~~\eta=0.65~\textrm{fm};~~U_0=40.5+0.13A~\textrm{MeV}$. $A$ is the mass number. As the mass $m$ of nucleon we took the quarter of the alpha particle mass, $m_\alpha=3727.379378~\textrm{MeV}/c^2$ ($\hbar c=197.3269804~\textrm{MeV}\cdot \textrm{fm}$).

It can be verified (as mentioned earlier) that for the example under consideration, the parameter $x_0$ is very small, in particular, $x_0<0.0012$.
So, keeping only terms of the first and zero degree in $x_0$, we obtain:
\begin{equation}
\label{e6}
F(\alpha,x_0)F(-\alpha,x_0)\underset{x_0\rightarrow 0}\simeq 1+\frac{2x_0\alpha^2}{1+4\alpha^2},
~~~~~~~~~
F(\alpha,x_0)^2\underset{x_0\rightarrow 0}\simeq
x_0^{2 i \alpha}\left(1+\frac{2x_0 \alpha^2}{1+2 i \alpha}\right).
\end{equation}
Additional use of the definition (\ref{w6}) yields:
\begin{eqnarray}
\label{e7}
F^2(-\alpha,x_0)J_1^{(S)}(x_0)+F^2(\alpha,x_0)J_2^{(S)}(x_0)-
2F(\alpha,x_0)F(-\alpha,x_0)J_3^{(S)}(x_0)\simeq
~~~~~~~~~~~~\nonumber~\\
\simeq 2\left\{\frac{R}{\eta}+\frac{2\alpha^2 \exp(-R/\eta)}{(1+4\alpha^2)^2}
\left[1-4\alpha^2+\frac{R}{\eta}(1+4\alpha^2)
\right]\right\}.~~~~~
\end{eqnarray}
Another feature of our example is that the parameter $\alpha$
is close to $1$ ($0.95<\alpha<1$).
Using the corresponding series expansion about $\alpha=1$, and limiting ourselves to the power of the fourth order, one obtains:
\begin{eqnarray}
\label{e8}
J_3^{(L)}-J_3^{(S)}(1)\underset{\alpha\rightarrow 1}{\simeq} C_3(\alpha)=
3.91601 0683 - 0.14477 63693 (\alpha-1)+
~~~~~~~~~~~~~~~~~\nonumber~\\
+0.15866 05205 (\alpha-1)^2 -
 0.13830 72935  (\alpha-1)^3 + 0.09449 89098 (\alpha-1)^4 .~~~~
\end{eqnarray}
\begin{eqnarray}
\label{e9}
J_1^{(L)}-J_1^{(S)}(1)\underset{\alpha\rightarrow 1}{\simeq} C_1(\alpha)=-0.9866370464 + 4.914762560~i-
~~~~~~~~~~~~~~~~~~~~~~~~~\nonumber~\\
 - (8.381381971- 6.402826365~i) (\alpha-1) -(14.74510685+4.544987922~i) (\alpha-1)^2-
~~\nonumber~\\
 - (2.941628785+15.20978607~i) (\alpha-1)^3 +(9.723029315 - 6.394376371~i) (\alpha-1)^4.~~~~
\end{eqnarray}
Eqs.(\ref{e8}) and (\ref{e9}) enable us to get rid of the onerous computations of infinite double power series.
Inserting representations (\ref{e6})-(\ref{e9}) into the RHS of Eq.(\ref{e5}), and using Eqs.(\ref{w6}) and (\ref{w20}), we obtain:
\begin{eqnarray}
\label{e10}
r_e=2R\left(1-\frac{R}{a}+\frac{R^2}{3a^2}\right)+4\eta N^2\left\{\frac{R}{\eta}-C_3+
\cos\left(\frac{2R\alpha}{\eta}\right)\textrm{Re}~ C_1-
\sin\left(\frac{2R\alpha}{\eta}\right)\textrm{Im}~ C_1+
\right.
~~~~\nonumber~\\
+\frac{2\alpha^2\exp(-R/\eta)}{1+4\alpha^2}
\left[\frac{R}{\eta}-C_3+\frac{1-4\alpha^2}{1+4\alpha^2}+
\left(\textrm{Re}~C_1-2 \alpha \textrm{Im}~C_1\right)\cos\left(\frac{2R\alpha}{\eta}\right)-
\right.
~~~~\nonumber~\\
\left.\left.
-\left(\textrm{Im}~C_1+2 \alpha \textrm{Re}~C_1\right)\sin\left(\frac{2R\alpha}{\eta}\right)
\right]\right\},~~~~~~~~~
\end{eqnarray}
where the scattering length $a$, and the normalization factor $N$ are defined by Eqs.(\ref{w10}) and (\ref{w13}), respectively.
Note that the functions $C_1(\alpha)$ and $C_3(\alpha)$ in the last equation have been replaced with $C_1$ and $C_3$, respectively, for ease of recording.

The scattering length $a$ and the effective range $r_e$ for the example \cite{PER,MJ1} mentioned above are
presented in Fig.\ref{F4} and \ref{F5}, respectively,  for $40\leq A\leq72$.
Remind that representation (\ref{e10}) for the effective range is suitable only for case when the parameter $x_0$ is small enough, and the parameter $\alpha$ is close to $1$.
However, one should emphasize that the relative error of the corresponding computations is less than $10^{-6}$ for the entire considered range of the mass number $A$.

It is worth noting that these figures demonstrate discontinuity points of the second kind at $A\simeq52.2504$ (for scattering length) and $A\simeq46.2708$ (for effective range), respectively.
This means that in the immediate vicinity of both points, the leading term of expansion (\ref{A0}) becomes $r_e k^2/2$.

All of the analytic results are fully coincident with the correspondent calculations carried out by
direct solution of the Schr$\ddot{\textrm{o}}$dinger equation (for $a$), or with the help of numerical integration (for $r_e$).

\section{Approximate analytical solution}\label{S3}

There are many potentials, such as, e. g., the Yukawa potential, that do not give exact analytic solutions to the Schr$\ddot{\textrm{o}}$dinger equation (\ref{A1}).
To calculate the scattering parameters for these potentials, we propose a technique that will be described below.

Using the definition (\ref{A4}) for the scattering length, let us present a solution of Eq.(\ref{A1}) in the form:
\begin{equation}
\label{A5}
\chi(r)= f(r)-1+\frac{r}{a},
\end{equation}
where according to the boundary conditions (\ref{A2}) and (\ref{B27a}) the function $f(r)$ must satisfy the conditions:
\begin{equation}
\label{A6}
f(0)=1,~~~~f(\infty)=0.
\end{equation}
This function, in turn, can be represented as an expansion in some basis $\left\lbrace g_n(r) \right\rbrace $:
\begin{equation}
\label{A7}
f(r)\simeq \sum_{n=0}^N C_n g_n(r).
\end{equation}
In this paper we will use the simplest well-known basis functions of the form
\begin{equation}
\label{A8}
g_n(r)=e^{-\beta r }r^n,~~~~~~~~~~~~~~~~~~~~~~~~~~~~(\beta>0)
\end{equation}
which enables us to satisfy automatically the second of conditions (\ref{A6}),
whereas one should put $C_0=1$ to satisfy the first one.

In this section, the system of units with $m=\hbar=1$ will be used.
In order to calculate the coefficients $C_i$, one can apply the well-known method of projecting the initial Eq.(\ref{A1}) onto the subspace of the basis functions (\ref{A8}).
Let the approximate solution $\chi(r)$ of this equation is presented by Eqs.(\ref{A5})-(\ref{A8}).
Multiplication of both sides of the equation (\ref{A1}) by the basis function $ g_k(r) $ followed by
integration over the whole space yields:
\begin{equation}
\label{B12} \int_{0}^{\infty}\chi''(r)g_k(r)dr=2\int_{0}^{\infty}V(r)\chi(r)g_k(r)dr
~~~~~~~~~~~~~~~(k=1,2,...,N).
\end{equation}
Substituting functions (\ref{A5}) with $f(r)$ of the form (\ref{A7}) into the equation (\ref{B12}),
one obtains the following system of linear equations for $C_n$:
\begin{equation}
\label{B13}
\sum_{n=1}^{N}C_nS_{nk}=\frac{q_k}{a}-h_k,
\end{equation}
where
\begin{equation}
\label{B14} S_{nk}=\int_{0}^{\infty}\left[ g_n''(r)-2V(r)g_n(r) \right]g_k(r)dr,
\end{equation}
\begin{equation}
\label{B15} q_{k}=2\int_{0}^{\infty}V(r)g_k(r)rdr,
\end{equation}
\begin{equation}
\label{B16} h_{k}=\int_{0}^{\infty} \left\lbrace 2V(r)\left[1-g_0(r)
\right]+g_0''(r) \right\rbrace g_k(r)dr.
\end{equation}
Solution of Eq.(\ref{B13}) can be written in the form
\begin{equation}
\label{B17}
C_n=\sum_{k=1}^N \left(  \frac{q_k}{a}+h_k\right)S_{kn}^{-1},
\end{equation}
where, $S_{kn}^{-1}$ are the elements of the inverse matrix in respect to the matrix with the elements $S_{nk}$.

Using the asymptotic behavior (\ref{A3}), it is easy to show, that the scattering length (\ref{A4}) can be presented in the form:
\begin{equation}
\label{A9}
a=\lim_{R\rightarrow \infty}\left[R-\frac{1}{y(R)} \right],
\end{equation}
where $y(R)=\chi'(R)/\chi(R)$ is the logarithmic derivative of the radial wave function $\chi(R)$.

At this stage we propose to apply the quasilinearization method (QLM) \cite{LEZ}, which enables us to calculate $y(R)$ using the analytic but very accurate approximation. The QLM is iterative one.
It was shown \cite{LEZ} that already at the first iteration the QLM can produce the analytical logarithmic derivative $y_1(R)$, which can be very accurate in case of making the correct choice of the wave function $\chi_0(R)$ of zero iteration (initial guess).
Thus, according to QLM the logarithmic derivative $y_1(R)=\chi_1'(R)/\chi_1(R)$ of the solution $\chi_1(R)$ to the Schr$\ddot{\textrm{o}}$dinger equation (\ref{A1}) can be presented in the form:
\begin{equation}
\label{A11}
y_1(R)=\frac{G(R)}{\chi_0^2(R)},
\end{equation}
where
\begin{equation}
\label{A12}
G(R)=\int_0^R\left[\chi_0'^2(r)+2V(r)\chi_0^2(r) \right]dr.
\end{equation}
Subsequent consideration is dependent on the choice of potential $V(r)$.


As an example, let us consider the Yukawa potential of the form:
\begin{equation}
\label{B1}
V(r)=-\lambda\frac{e^{-r}}{r}.~~~~~~~~~~~~~~~~~~~~~(\lambda>0)
\end{equation}
This potential, also called the "screened Coulomb potential", is used in various fields of physics to model singular but short-range interactions.
Note, that the scale transformation \cite{PAT} for the scattering length
\begin{equation}
\label{B2}
a(\lambda,r_0)=r_0a(\lambda r_0,1)
\end{equation}
enables us to investigate the Yukawa potential of the simplified form (\ref{B1}) instead of the general form corresponding to substitution of the exponent $-r$ by $-r/r_0$.

Thus, using the function defined by Eqs.(\ref{A5})-(\ref{A8}) as the initial guess $\chi_0(r)$, one obtains for the integral (\ref{A12}) with the Yukawa potential (\ref{B1}):
\begin{eqnarray}
\label{B3}
G(R)=\frac{R}{a^2}-b_0+b_1(R)e^{-R}+b_2(R)e^{-R \beta}+b_3e^{-R(1+\beta)}-\frac{\beta}{2}e^{-2R\beta}+
~~~~~~~~~~~~~~~~~~~~~~~~~~~~~~~~~~~~~\nonumber~\\
+2\lambda\left\lbrace
\Gamma(0,R)-2\Gamma[0,R(1+\beta)]+\Gamma[0,R(1+2\beta)]
\right\rbrace +
~~~~~~~~~~~~~~~~~~~~~~~~~~~~~~~~~~~~~~\nonumber~\\
+\sum_{n=1}^NC_n\left\lbrace
-p_{0,n}+p_{1,n}(R)e^{-2R\beta}+p_{2,n}(R)e^{-R(1+\beta)}+p_{3,n}\Gamma(n,2R\beta)+
p_{4,n}\Gamma[n,R(1+2\beta)]+
\right.
~~~~~~~~~~~~~~~~\nonumber~\\
\left.
+p_{5,n} \Gamma[n,R(1+\beta)]
\right\rbrace
-\sum_{n=1}^N\sum_{k=1}^NC_nC_k\left\lbrace
s_{0,n k}+s_{1,n k}\Gamma[n+k,R(1+2\beta)]+s_{2,n k}\Gamma[n+k-1,2R\beta]+
\right.
~~~~~~~~~\nonumber~\\
\left.
+s_{3,n k}\Gamma[n+k,2R\beta]+s_{4,n k}\Gamma[n+k+1,2R\beta]
\right\rbrace,~~~~~~~~~~~~~~~~~~~~~~~~~~
\end{eqnarray}
where
\begin{eqnarray}
\label{B3a}
b_0=\frac{2\lambda}{a^2}-\frac{2}{a}\left(2\lambda-1-\frac{2\lambda}{1+\beta} \right)
-\frac{\beta}{2}+2\lambda
\ln \left[ \frac{(1+\beta)^2}{1+2\beta}\right];
~~~~~~~~~~~~~~~~~~~~~~~~~~\nonumber~\\
p_{0,n}=(n-1)!\left\lbrace  \frac{n \beta}{(2\beta)^n}+
4\lambda\left[\frac{1}{(1+2\beta)^n}+\frac{n-a(1+\beta)}{a(1+\beta)^{n+1}} \right]
\right\rbrace ;
~~~~~~~~~~~~~~~~~~~~~~~~~~\nonumber~\\
s_{0,nk}=\frac{2\lambda(n+k-1)!}{(1+2\beta)^{n+k}}+
\frac{\left[(n-k)^2-n-k \right](n+k-2)! }{4(2\beta)^{n+k-1}};
~~~~~~~~~~~~~~~~~~~~~~~~~~~~~~~~~~~
\end{eqnarray}
\begin{eqnarray}
\label{B4}
b_1(R)=\frac{2\lambda(1-2a+R)}{a^2};~~
b_2(R)=\frac{2}{a}\left(1+\sum_{n=1}^NC_nR^n \right);~~
b_3=\frac{4\lambda}{a(1+\beta)};~~
p_{1,n}(R)=-\beta R^n;
~~\nonumber~\\
p_{2,n}(R)=\frac{4\lambda R^n}{a(1+\beta)};~~
p_{3,n}=\frac{n \beta}{(2\beta)^n};~~
p_{4,n}=\frac{4\lambda}{(1+2\beta)^n};~~
p_{5,n}=\frac{4\lambda\left[n-a(1+\beta) \right] }{a(1+\beta)^{n+1}};
~~~~~~~~~~~~~~~~~~\nonumber~\\
s_{1,nk}=-\frac{2\lambda}{(1+2\beta)^{n+k}};~~
s_{2,nk}=\frac{n k}{(2\beta)^{n+k-1}};~~
s_{3,nk}=-\frac{(n+k)}{2(2\beta)^{n+k-1}};~~
s_{4,nk}=\frac{1}{4(2\beta)^{n+k-1}}.
~~~~~~~~~~~~~
\end{eqnarray}
Taking into account, that the incomplete gamma functions $\Gamma(x,y)$
for integer $x=n$ can be presented in the form
\begin{eqnarray}
\label{B5}
\Gamma(n,x)=(n-1)!e^{-x}\sum_{k=0}^{n-1}\frac{x^k}{k!},~~~~~~~~~(n=1,2,...)
~~~~~~~~~~~~~~~~~~~~~~~~~\nonumber~\\
\Gamma(0,x)=\frac{e^{-x}}{x+1-\frac{1}{x+3-...}},~~~~~~~~~~~~~~~~~~~~~~~~~~~~~~~~~~~~~~~~~~~~~~~~~~~~~~~~~~~~~~
\end{eqnarray}
one can write the following asymptotic expression for the integral (\ref{A12}):
\begin{equation}
\label{B6}
G(R)\underset{R\rightarrow\infty}{\simeq} \frac{R}{a^2}-b_0-\sum_{n=1}^NC_np_{0,n}-\sum_{n=1}^N\sum_{k=1}^NC_nC_ks_{0,nk}.
\end{equation}
To obtain the last equation, the terms with exponential factors  were neglected.

It is clear, that the asymptotic expression for the initial guess function (\ref{A5}) is of the form:
\begin{equation}
\label{B7}
\chi_0(R)\underset{R\rightarrow\infty}{\simeq}\frac{R}{a}-1.
\end{equation}
Substituting the asymptotic expressions (\ref{B6}) and (\ref{B7}) into the RHS of Eq.(\ref{A11}), one obtains for the RHS of Eq.(\ref{A9}):
\begin{equation}
\label{B8}
\lim_{R\rightarrow\infty}\left[R-\frac{\chi_0^2(R)}{G(R)} \right]=
a\left[2-a\left(
b_0+\sum_{n=1}^NC_np_{0,n}+\sum_{n=1}^N\sum_{k=1}^NC_nC_k s_{0,n k}
\right)
\right].
\end{equation}
Thus, Eq.(\ref{A9}) can be rewritten in the form:
\begin{equation}
\label{B9}
 b_0+\sum_{n=1}^NC_n p_{0,n}+\sum_{n=1}^N\sum_{k=1}^NC_nC_ks_{0,nk}=\frac{1}{a}.
\end{equation}
Inserting the explicit expressions (\ref{B3a}) for $b_0,p_{0,n}$ and $s_{0,nk}$
into Eq.(\ref{B9}), one obtains the equation
\begin{equation}
\label{B10}
A_2 a^2+A_1 a +2\lambda=0,
\end{equation}
where
\begin{eqnarray}
\label{B11}
A_1=1-\frac{4\lambda}{1+\beta}\left[
\beta-\sum_{n=1}^N\frac{C_n n(n-1)!}{(1+\beta)^n}\right],
~~~~~~~~~~~~~~~~~~~~~~~~~~~~~~~~~\nonumber~\\
A_2=-\frac{\beta}{2}+2\lambda \ln\left[\frac{(1+\beta)^2}{1+2\beta} \right]+
\sum_{n=1}^NC_n(n-1)!\left\lbrace
\frac{n\beta}{(2\beta)^n}+4\lambda\left[
\frac{1}{(1+2\beta)^n}-\frac{1}{(1+\beta)^n}
\right]
\right\rbrace+
~~~~~~~~~\nonumber~\\
+\sum_{n=1}^N \sum_{k=1}^N C_n C_k(n+k-2)!\left\lbrace
\frac{2\lambda(n+k-1)}{(1+2\beta)^{n+k}}-
\frac{\beta \left[n+k-(n-k)^2 \right] }{2(2\beta)^{n+k}}
\right\rbrace. ~~~~~~~~~~~
\end{eqnarray}
In fact, Eq.(\ref{B10}) represents a transcendental equation for the scattering length $a$, because in general case both linear coefficients $C_i$ and the exponent factor $\beta$ can be functions of $a$.
However, in the case of $\beta$ is independent on $a$, one obtains a quadratic equation for $a$, and hence, an explicit analytic expression for the scattering length.
Let us demonstrate this point on examples of the simplest expansions with $N=1,2$.

In order to derive an analytic solutions to Eq.(\ref{B13}) corresponding to the Yukawa potential (\ref{B1}), let us write the explicit expressions for $S_{nk},q_k$ and $h_k$ according to Eqs.(\ref{B14}), (\ref{B15}) and (\ref{B16}), respectively:
\begin{eqnarray}
\label{B20}
S_{nk}=(n+k-2)!\left\lbrace
\frac{2\lambda(n+k-1)}{(1+2\beta)^{n+k}}-\frac{\beta\left[
n+k-(n-k)^2\right] }{2(2\beta)^{n+k}}
\right\rbrace,
~~~~~~~~~~~~~~~~~~~~~~~~~~~~~\nonumber~\\
q_k=-\frac{2\lambda k!}{(1+\beta)^{1+k}},
~~~~~~~~~~~~~~~~~~~~~~~~~~~~~~~~~~~~~~~~~~~~~~~~~~~~~~~~~~~~~~~~~~~~~~~~~~~~~~~~~~~ \nonumber~\\
h_k=(k-1)!\left\lbrace
2\lambda\left[
\frac{1}{(1+2\beta)^k}-\frac{1}{(1+\beta)^k}\right]
+\frac{k \beta}{2(2\beta)^k}
\right\rbrace .~~~~~~~~~~~~~~~~~~~~~~~~~~~~~~~~~~~~~~
\end{eqnarray}
Thus, solution of Eq.(\ref{B13}) for the two-term expansion ($N=1$) yields:
\begin{equation}
\label{B21}
C_1^{(1)}=\frac{\beta(1+2\beta)\left\lbrace
a(1+\beta)\left[ 1+\beta(3+2\beta-8\lambda)\right]+8\lambda(1+2\beta)
\right\rbrace}
{a(1+\beta)^2\left[ 1+4\beta(1+\beta-2\lambda)\right] }.
\end{equation}
The superscript denotes that this expression corresponds to the expansion with $N=1$.
Inserting this expression into Eqs.(\ref{B11}), one obtains the following simple equation, instead of Eq.(\ref{B10}):
\begin{equation}
\label{B22}
B_2^{(1)}a^2+B_1^{(1)}a+B_0^{(1)}=0,
\end{equation}
with the coefficients:
\begin{eqnarray}
\label{B23}
B_2^{(1)}=-\frac{1}{4}\left[
\beta(1+\beta)^2(1+2\beta)^2+16\lambda\beta^3(1+\beta)-64\lambda^2\beta^3
\right]+
~~~~~~~~~~~~~~~~~~~~~~~~~~~~~~~~~~~~~~~~~~~~~\nonumber~\\
2\lambda(1+\beta)^2\left[1+4\beta(1+\beta-2\lambda) \right]
\ln\left[ \frac{(1+\beta)^2}{1+2\beta}\right],
~~~~~~~~~~~~~~~~~~~~~~~~~~~~~~~\nonumber~\\
B_1^{(1)}=(1+\beta)^2(1+2\beta)^2-4\lambda\beta\left[2+\beta(5+6\beta+4\beta^2) \right]
+\dfrac{32\lambda^2\beta^4}{1+\beta},
~~~~~~~~~~~~~~~~~~~~~~~~~~~~~~~~~~~~~~~\nonumber~\\
B_0^{(1)}=\frac{2\lambda}{(1+\beta)^{2}}\left\lbrace
(1+\beta)^4(1+2\beta)^2-8\lambda\beta^3\left[2(1+2\beta)+\beta^2 \right]
\right\rbrace.
~~~~~~~~~~~~~~~~~~~~~~~~~~~~~~~~~~~~~~~~~~~~~
\end{eqnarray}
Solving Eq.(\ref{B13}) for the three-term expansion ($N=2$), one can derive the proper expressions for the linear coefficients $C_1^{(2)}$ and $C_2^{(2)}$. Inserting then those expressions into Eqs.(\ref{B11}), one can obtain the equation like (\ref{B22}), but with coefficients $B_k^{(2)}$ instead of $B_k^{(1)}$.
Note, that the coefficients $B_k^{(N)}$ ($k=0,1,2$) depend only on parameters $\lambda$ and $\beta$.
Therefore, if the exponent $\beta$ is independent on $a$, then Eq.(\ref{B22}) represents a simple quadratic equation for the scattering length $a$.

The simplest representation for the exponent $\beta$, depending only on the parameter $\lambda$, can be obtained as follows. The basic Eq.(\ref{A1}) for the Yukawa potential (\ref{B1}) at small $r$ can be written in the form:
\begin{equation}
\label{B24}
\chi''(r)+\frac{2\lambda(1-r)}{r}\chi(r)=0.
\end{equation}
The general solution of this equation is:
\begin{equation}
\label{B25}
\chi(r)=e^{-r\sqrt{2\lambda}}r\left[
c_1~_1F_1\left( 1-\sqrt{\frac{\lambda}{2}},2,2r\sqrt{2\lambda}\right) +
c_2~U\left( 1-\sqrt{\frac{\lambda}{2}},2,2r\sqrt{2\lambda}\right)
\right],
\end{equation}
 where $_1F_1(n,m,z)$ and $U(n,m,z)$ are the confluent hypergeometric functions of the first and second kind, respectively.
Solution of the form (\ref{B25}) puts some ideas on trying $\beta=\sqrt{2\lambda}$.
In this case, the only solution (with positive square root) of the quadratic equation (\ref{B22})
\begin{equation}
\label{B26}
a=\frac{-B_1^{(N)}+\sqrt{\left( B_1^{(N)}\right)^2-4B_0^{(N)}B_2^{(N)} }}{2B_2^{(N)}}
\end{equation}
is correct for arbitrary $N$.

Thus, substituting the representation (\ref{A5})-(\ref{A8}) into the RHS of Eq.(\ref{B27}), one obtains the following simple expression for the effective range:
\begin{equation}
\label{B28}
r_e=\frac{4}{\beta}\left[
\sum_{n=0}^NC_n\frac{n!}{\beta^{n}}\left( 1-\frac{n+1}{a \beta}\right)-
\frac{1}{4}\sum_{n=0}^N\sum_{k=0}^NC_nC_k\frac{(n+k)!}{(2\beta)^{n+k}}
\right].
\end{equation}
The results of computations of the scattering length $a$ with using the exponent $\beta=\sqrt{2\lambda}$ are presented in the Table \ref{T1} for the expansion length (basis size) $N$ from $1$ to $7$.
The "exact" values obtained by the direct numerical integration of Eq.(\ref{A1}) are presented for comparison, as well. These values coincide completely
with the results for the scattering length presented in Ref. \cite{Hor}.
An exclamation mark (in the Table \ref{T1}) at the number denotes that the corresponding value of $a$ has an imaginary part, and therefore, only $\textrm{Re}(a)$ is presented in the Table.
It is seen, that the accuracy of the approximate results increases with the expansion length $N$, as it was expected.
The accuracy decreases with increasing the coupling constant $\lambda$.
However, the results in Table \ref{T1} demonstrate that for $\lambda\leqslant 3.5$
the expansion length $N\leqslant 7$ is enough to provide the scattering length accuracy of the same order as the "exact" data presented ibidem.
To provide the "exact" accuracy for $a$ corresponding to the Yukawa potential (\ref{B1}) with the coupling constant $\lambda>3.5$ one needs a greater $N$. In particular, according to our calculations at least the expansion lengths $N=11,14$ and $15$ give the "exact" accuracy for $\lambda=4.5,5.5$ and $6.5$, respectively.
On the other hand, even the shortest expansion with $N=1$ gives a good results for small $\lambda$.
The larger $\lambda$ requires the larger $N$.

In Table \ref{T1} the exact effective range $r_e^{exact}$ obtained with
using the "exact" numerical wave functions $\chi(r)$ is presented. The minimal expansion length
$N_{min}$, which provides the "exact" accuracy for $r_e$ (according to
Eq.(\ref{B28})) is presented in the last column of the Table \ref{T1}. It is seen that the
values of $N_{min}$ for $r_e$ are greater, as a rule, than the ones for the
scattering length $a$.

\section{Conclusions}\label{S4}

The properties of the effective-range expansion have been studied.
The analytic expressions for the leading terms of this expansion, the scattering length and effective range, have been derived for the inverse-power potential and the Woods-Saxon potential.

The plots for the scattering length producing by the inverse-power potential of the form $\alpha r^{-\beta}$ with parameters $\beta>3$ and $\alpha=0.1, 1, 10$, are presented in Fig. \ref{F1} in the units of $\hbar^2/2m$.
The corresponding effective range  is shown in Fig. \ref{F2}
for the same values of $\alpha$ but for $3<\beta<6$, whereas Fig. \ref{F3} corresponds to $\beta>6$.
Because of the different scales, the graphs for $3<\beta<6$ and $\beta>6$ are presented on individual figures, which demonstrate a few interesting features of the effective range.
First, two points,  $\beta=10/3$ and $\beta=6$, of zero effective range have been found.
Second, there have been revealed three $\beta$-intervals, $[0,10/3],[3.5,4],[5,6]$, for negative effective range, and three $\beta$-intervals, $[10/3,3.5],[4,5],[6,\infty]$, for positive one.
Third, four $\beta$-points of discontinuity of the second kind have been found
for  $\beta=3,3.5,4,5$.

\begin{figure}
\caption{Scattering length for the Woods-Saxon potential presenting the real part of the optical-model calculations for the mass number $40\leq A \leq 72$.}
\includegraphics[width=5.5in]{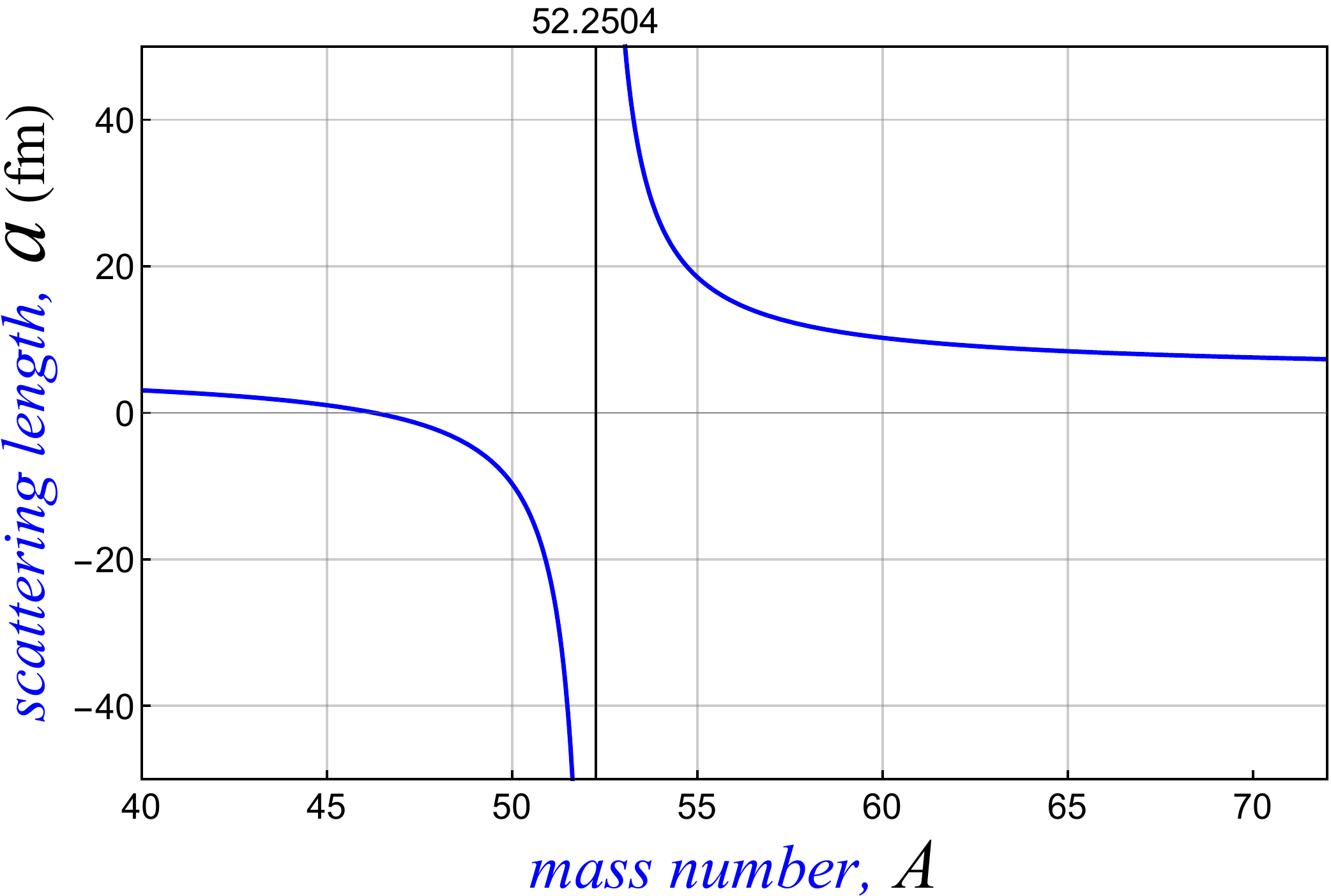}
\label{F4}
\end{figure}

\begin{figure}
\caption{Effective range $r_e$ for the Woods-Saxon potential presenting the real part of the optical-model calculations for the mass number $40\leq A \leq 72$.}
\includegraphics[width=5.5in]{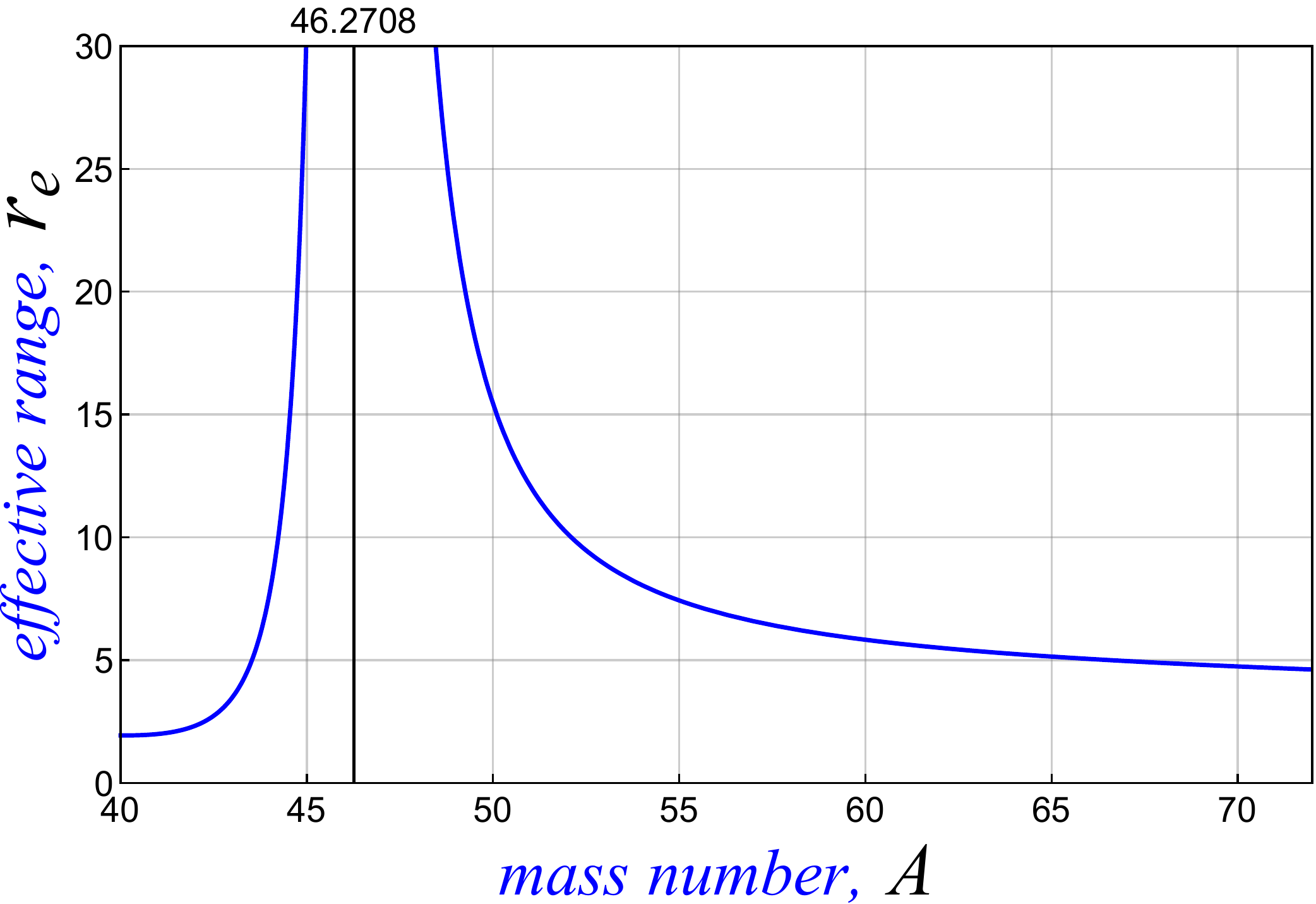}
\label{F5}
\end{figure}

The scattering parameters for the Woods-Saxon potential applied to the optical-model calculations \cite{MJ2} are presented in Figs.\ref{F4} and \ref{F5}  for the mass number $40\leq A\leq72$.
Note that the simplified expression (see Eq.(\ref{e10})) for the effective range was derived  only for the case of a small parameter $x_0$, and a parameter $\alpha$ close to $1$.
However, one should emphasize that the relative error of the corresponding computations is less than $10^{-6}$. 
It is worth noting that figures \ref{F4} and \ref{F5} demonstrate discontinuity points of the second kind at $A\simeq52.2504$ (for scattering length) and $A\simeq46.2708$ (for effective range), respectively,
which means that in the immediate vicinity of both points, the leading term of expansion (\ref{A0}) becomes $r_e k^2/2$.

A technique was developed for the approximate calculation of the scattering parameters.
Approximate analytic formulas representing the scattering length and effective range have been obtained for the Yukawa potential as the example. The corresponding results are presented in Table \ref{T1} for  different values of the coupling constant $\lambda$ in the range $[0.1,6.5]$.

All analytic formulas, both exact and approximate, have been verified by comparing with the corresponding results obtained by direct numerical calculations.
The presented results can be used with advantage in the fields of nuclear physics,
atomic and molecular physics, quantum chemistry and others.

\section{Acknowledgment}\label{S5}

I acknowledge helpful discussions with Prof. N. Barnea.


\newpage

\begin{table}
\begin{center}
\caption{The scattering length $a$ and the effective range $r_e$ for the Yukawa
potential $V(r)=-\lambda e^{-r}/r$. The exponent $\beta=\sqrt{2\lambda}$ is
taken for representation of the basis functions (\ref{A8}). $N_{min}$ denotes the
minimal expansion length $N$ providing the values of $r_e=r_e^{exact}$.}
\begin{tabular}{|c||c|c|c|c|c|c|c||c||c|c|}
\hline {\large $\lambda\diagdown N$}& 1& 2& 3& 4& 5& 6& 7& $a^{exact}$&
$r_{e}^{exact}$& $N_{min}$\tabularnewline \hline \hline {\footnotesize 0.1}&
{\footnotesize -0.221844}& {\footnotesize -0.222487}& {\footnotesize -0.222588}&
{\footnotesize -0.222606}& {\footnotesize -0.222609}& {\footnotesize -0.222610}&
{\footnotesize - " -}& {\footnotesize -0.222610}& {\footnotesize 20.12}&
{\footnotesize 27}\tabularnewline \hline {\footnotesize 0.5}& {\footnotesize
-2.20432}& {\footnotesize -2.20668}& {\footnotesize -2.20689}& {\footnotesize
-2.20691}& {\footnotesize -2.20692}& {\footnotesize - " -}& {\footnotesize - " -}&
{\footnotesize -2.20692}& {\footnotesize 3.934}& {\footnotesize 11}\tabularnewline
\hline {\footnotesize 1.}& {\footnotesize 7.93367}& {\footnotesize 7.91309}&
{\footnotesize 7.91150}& {\footnotesize 7.91138}& {\footnotesize - " -}&
{\footnotesize - " -}& {\footnotesize - " -}& {\footnotesize 7.91138}&
{\footnotesize 1.637}& {\footnotesize 7}\tabularnewline \hline {\footnotesize 1.5}&
{\footnotesize 2.14416}& {\footnotesize 2.13262}& {\footnotesize 2.12846}&
{\footnotesize 2.12843}& {\footnotesize 2.12841}& {\footnotesize - " -}&
{\footnotesize - " -}& {\footnotesize 2.12841}& {\footnotesize 0.7406}&
{\footnotesize 9}\tabularnewline \hline {\footnotesize 2.5}& {\footnotesize
-0.79336}& {\footnotesize -1.02589}& {\footnotesize -1.10746}& {\footnotesize
-1.11457}& {\footnotesize 1.11598}& {\footnotesize -1.11613}& {\footnotesize
-1.11616}& {\footnotesize -1.11616}& {\footnotesize 21.44}& {\footnotesize
13}\tabularnewline \hline {\footnotesize 3.5}& {\footnotesize 12.47262}&
{\footnotesize 11.00317}& {\footnotesize 10.55154}& {\footnotesize 10.53104}&
{\footnotesize 10.52685}& {\footnotesize 10.52661}& {\footnotesize 10.52658}&
{\footnotesize 10.52658}& {\footnotesize 3.011}& {\footnotesize 9}\tabularnewline
\hline {\footnotesize 4.5}& {\footnotesize 2.86636}& {\footnotesize 2.74 !}&
{\footnotesize 2.95 !}& {\footnotesize 2.98514}& {\footnotesize 2.95399}&
{\footnotesize 2.94821}& {\footnotesize 2.94694}& {\footnotesize 2.94651}&
{\footnotesize 1.447}& {\footnotesize 18}\tabularnewline \hline {\footnotesize 5.5}&
{\footnotesize 2.08 !}& {\footnotesize 1.99 !}& {\footnotesize 1.40082}&
{\footnotesize 1.14152}& {\footnotesize 1.06117 }& {\footnotesize 1.03526}&
{\footnotesize 1.02778}& {\footnotesize 1.02472}& {\footnotesize 6.622}&
{\footnotesize 26}\tabularnewline \hline {\footnotesize 6.5}& {\footnotesize 1.83
!}& {\footnotesize 0.472875}& {\footnotesize -1.01017}& {\footnotesize -2.20354}&
{\footnotesize -2.68053}& {\footnotesize -2.85253}& {\footnotesize -2.90604}&
{\footnotesize -2.92986}& {\footnotesize 13.10}& {\footnotesize 21}\tabularnewline
\hline
\end{tabular}
\label{T1}
\end{center}
\end{table}

\end{document}